\documentclass{iau}

\usepackage{amsmath}
\usepackage{graphicx}
\usepackage{multirow}

\begin{document}

\lefttitle{Parker et al.}
\righttitle{Science from PNe in the Galactic Bulge}

\jnlDoiYr{2023}
\doival{10.1017/xxxxx}
\volno{384}

\aopheadtitle{Proceedings IAU Symposium}
\editors{O. De Marco, A. Zijlstra, R. Szczerba, eds.}

\title{Diverse Science from VLT imagery and spectroscopy of PNe in the Galactic Bulge}

\author{Quentin Parker$^{1}$, Shuyu Tan$^{1}$, Andreas Ritter$^{1}$, Albert Zijlstra$^{2}$ }
\affiliation{1: Laboratory for space research, University of Hong Kong, HK SAR PRC; 2: Manchester University, U.K.}
%\affiliation{Taiyuan University, PRC}

\begin{abstract}
We have undertaken a deep investigation of a well defined sample of 136 PNe located in a $10\times10$ 
degree central region of the Galactic Bulge observed with the ESO VLT and supplemented by archival HST 
imagery. These studies have provided precise morphologies, major axes position angles and the most 
robust sample of consistently derived chemical abundances available to date. Using these data we have 
statistically confirmed, at 5$\sigma$, the precise PNe population that provides the PNe alignment of major 
axes previously suggested in the Galactic Bulge, revealed a partial solution to the sulfur anomaly and 
uncovered interesting morphological, abundance and kinematic features. We summarise the most significant 
findings here with detailed results appearing in a series of related publications.
\end{abstract}


\maketitle

\section{Introduction}

There are currently $\sim3800$ Galactic planetary nebulae (PNe) known as recorded in the HASH database, 
e.g. \cite{Parker2016} with half arising from discoveries over the last 25 years by our group, e.g. 
\cite{Parker2006}, \cite{Miszalski2008}, \cite{Sabin2014}. These PNe have been found in crowded, 
wide-field, narrow-band H$\alpha$ surveys in the Galactic plane including the Galactic Bulge. Here we 
present summary results from our ESO VLT narrow-band imagery for
PNe located in the inner 10~$\times$~10 degree region of the Galactic bulge selected to be highly likely 
physical bulge members. This is based on the criteria detailed in \cite{Rees2013}. 
The 136 original targets were spectroscopically obtained using FORS2 on the ESO 8.2m VLT 
between 2015 and 2019 for PIs Rees, Zijlstra and Parker using the same instrumental 
configurations and observing strategies. We believe this represents the most self-consistent 
and well characterised sample of PNe available with high S/N spectra, from which 
well-determined spectroscopic properties and deep analysis can be performed.  
FORS2 narrow-band pre-imaging was also obtained and supplemented by 40 objects 
previously observed with \emph{HST}. These data have provided the basis for 
several detailed investigations either recently published, in press or about to 
be submitted. Selected scientific results that emerged are summarised conveniently below.

\section{The bulge PNe morphology, central stars and kinematic ages project}
Our high quality VLT pre-imaging and, when available, HST imagery, shows that a large fraction (68\%) of 
our compact bulge sample are bipolars \cite{Tan2023a}. The increase in the fraction of bipolar PNe 
comprising our bulge sample is very significant for a population ostensibly dominated by older stars. If 
bipolars generally derive from higher mass progenitors, as inferred by their preponderance at lower 
Galactic scale heights in the Galactic disk, as well as being in binaries, 
then that would suggest a significant, young stellar population in the bulge 
feeding the observed population today. They dominate all PNe 
emerging from the general, several Giga year old bulge stellar population.

\begin{figure}
\begin{center} 
  \includegraphics[scale=0.6]{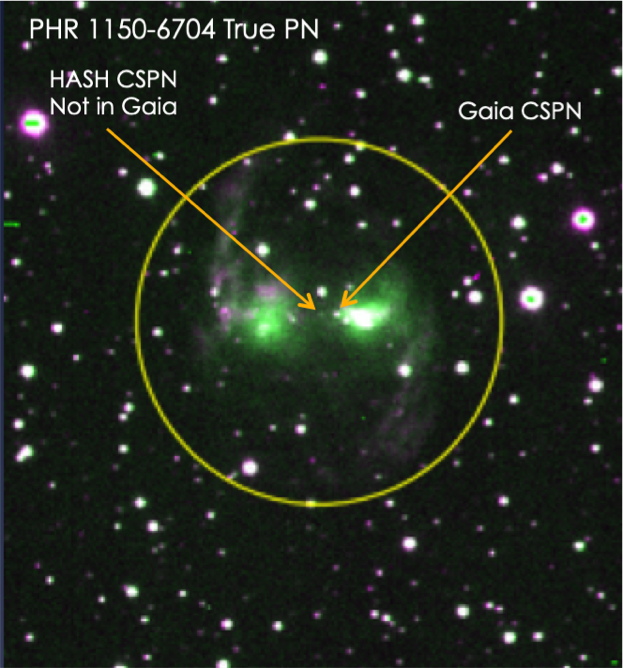}
  \caption{Example of a PNe (PHR1150-6704, HASH \#~2722) where the correct CSPN is clear in the HASH 
  imagery but has been incorrectly identified from Gaia data.}
  \label{cspn}
  \end{center}
\end{figure}

Four new planetary nebula central stars are also identified which are not in \emph{Gaia}. 
Some 11 putative central stars previously reported, based on \emph{Gaia} positions, 
are now shown not to be the true central star, refer Figure~\ref{cspn} for an example. 
Furthermore, \emph{Gaia} central stars reported in the literature are often based on the 
overall centroid position of a very compact planetary nebula rather than the actual 
central star within it that is often below Gaia limits. We find that 15\% of \emph{Gaia} CSPN identified 
by \citep{CW2021} are not confirmed. Great care is needed when using \emph{Gaia} CSPN compilations and we 
strongly recommend using the HASH database\footnote{HASH: available online 
\url{http://www.hashpn.space}} \cite{Parker2016} for such studies where every CSPN listed has been 
holistically vetted. HASH consolidates and federates available multi-wavelength imaging, spectroscopic 
and other data for Galactic PNe.  

\begin{figure}
\begin{center} 
  \includegraphics[scale=0.6]{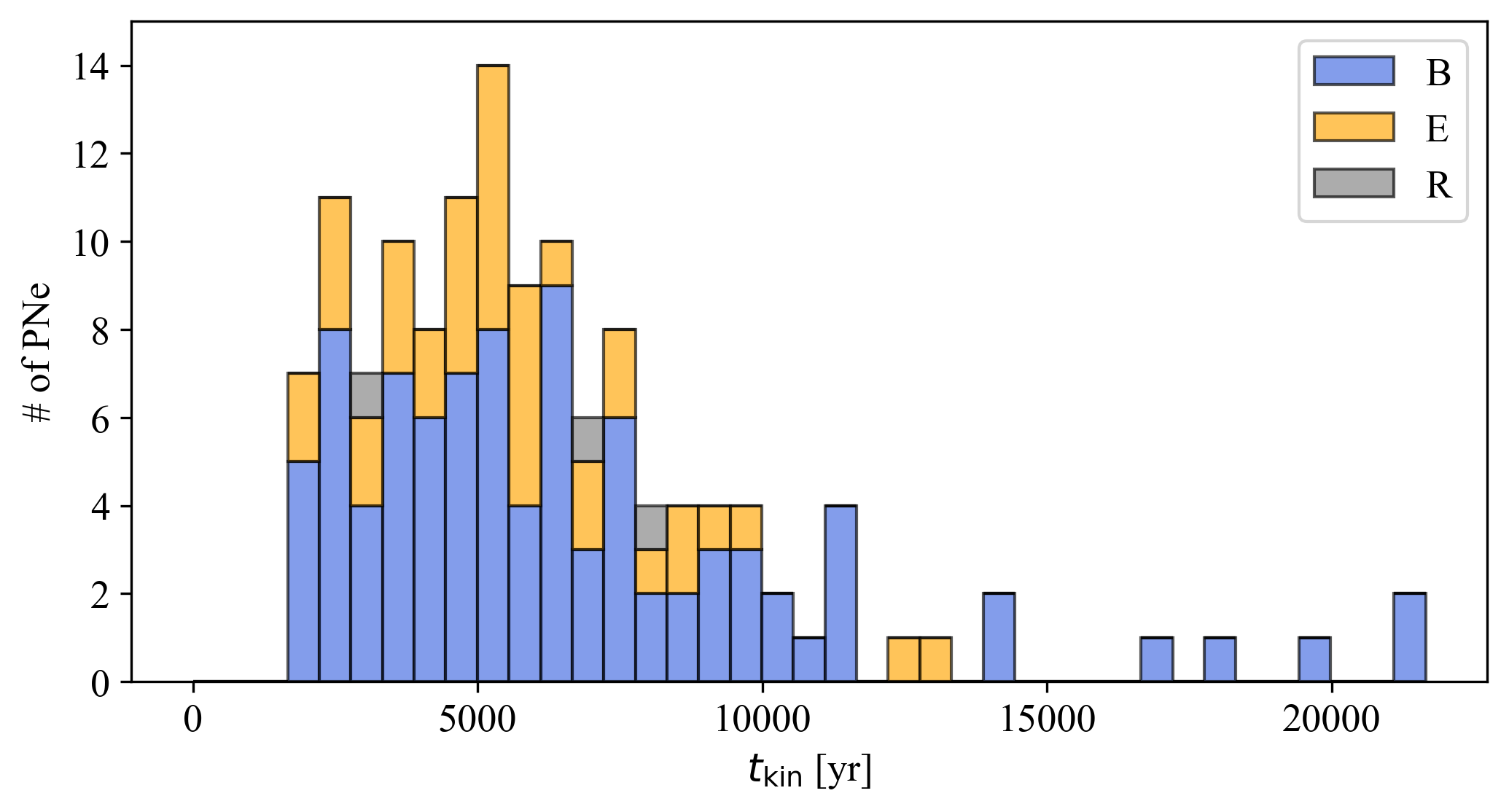}
  \caption{Distribution of kinematic ages determined for our bulge PNe.
Histograms are color-coded according to the main morphological classes.}
  \label{kinematic-age}
  \end{center}
\end{figure}
We also provided in \cite{Tan2023a} the most complete and accurate kinematic age estimates for our 
compact bulge PNe ever compiled. This included \emph{Gaia} distances for 33, 
accurate expansion velocities for 77, improved angular size determinations for all 
(from the HST, VLT and Pan-STARRS imagery) and average distance and expansion 
velocities for the rest. The expected young age of the PNe sample is confirmed 
with an average kinematic age of 6,400~$\pm$~3,800 years. Kinematically older 
PNe are dominated by bipolars with 14/16 PNe with estimated ages $>$10,000 years 
being bipolars (refer Figure~\ref{kinematic-age}). The best current work 
also puts the short-period binary fraction of 
bulge PNe at 15-20\% - maximum 60\% in bipolars \citep{Miszalski2009}, giving 9-12\% of PNe in 
binaries are bipolars. This is a factor of up to 7 lower than the bipolar fraction we find. A younger 
stellar population in the bulge, wider period binaries and the relationship between PN formation and the 
CSPN binarity may play a role but the issue of the progenitor masses of bipolars remains. 

\section{The bulge PNe major axes alignment project}
\begin{figure}
\begin{center} 
  \includegraphics[scale=.2]{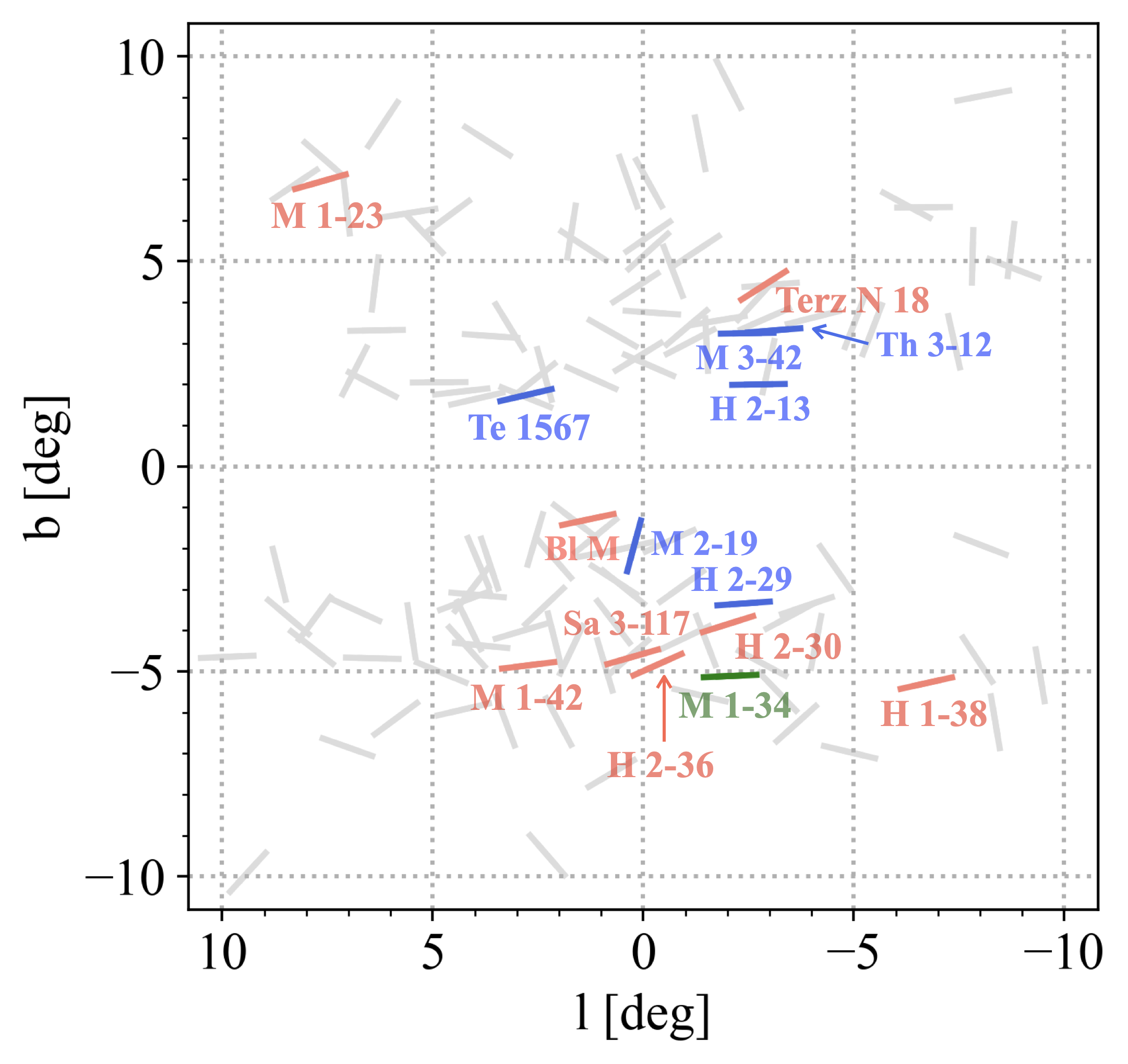}
  \caption{Distribution of measured GPAs for our bulge PNe. Grey vectors are
    the general sample, those with confirmed binary central stars are blue vectors and
    those with high abundance discrepancy factors, \textit{adfs} (that are taken as a reliable proxy for PNe hosting
    short-period binaries) are pink vectors. M~1-34, suspected of hosting a short period binary, is 
    plotted with a green vector.}
  \label{GPA-distribution}
  \end{center}
\end{figure}

We have recently reported a 5$\sigma$ PNe major axes alignment, almost parallel 
to the Galactic plane, for a subset of bulge PNe that host, or are inferred to host, 
short period binaries \citep{Tan2023b}.  Almost all are bipolar. It is only this 
very particular PNe sub-population that gives this result compared to the much weaker 
statistical alignments reported previously for PNe in the bulge where this signal is 
diluted (refer Figure~\ref{GPA-distribution}). 
We now have clear, irrefutable evidence of a organised and long lived 
process at play at the centre of our Galaxy for this particular PNe subgroup. 
Not only that but it has to have endured over cosmologically significant time periods 
given the age differences estimated for several PNe in the sample \citep{Tan2023b} 
that emerge from stars of different progenitor mass. The only plausible mechanism 
that could affect multiple close binary star orbits across large volumes of the Galaxy, 
as revealed by the observed major axes orientations of their eventual PNe, is thought 
to be magnetic fields. We speculate that PNe forming from wider binaries could have 
their symmetry axes change over time more randomly due to the ambient field where 
directions may evolve due to Galactic wind or bulk ISM motions. This provides a 
possible explanation for the observed alignment only being seen in post-CE binary PNe. 
If this is the process then it must remain ordered,  potent and have endured for 
billions of years. The only alternative is a pervasive force (magnetic?) acting 
currently only on the PNe bipolar lobe ejection mechanism that host short-period 
binaries. This must be sufficiently strong to align Galactic Position Angles (GPAs, see \citep{Tan2023b} for definition) over entire Galactic bulge 
in only a few thousand years which seems unlikely as current Galactic magnetic field 
strengths are too weak.

\section{The bulge PNe abundance project}
\begin{figure}
  \begin{center} 
    \includegraphics[scale=.8]{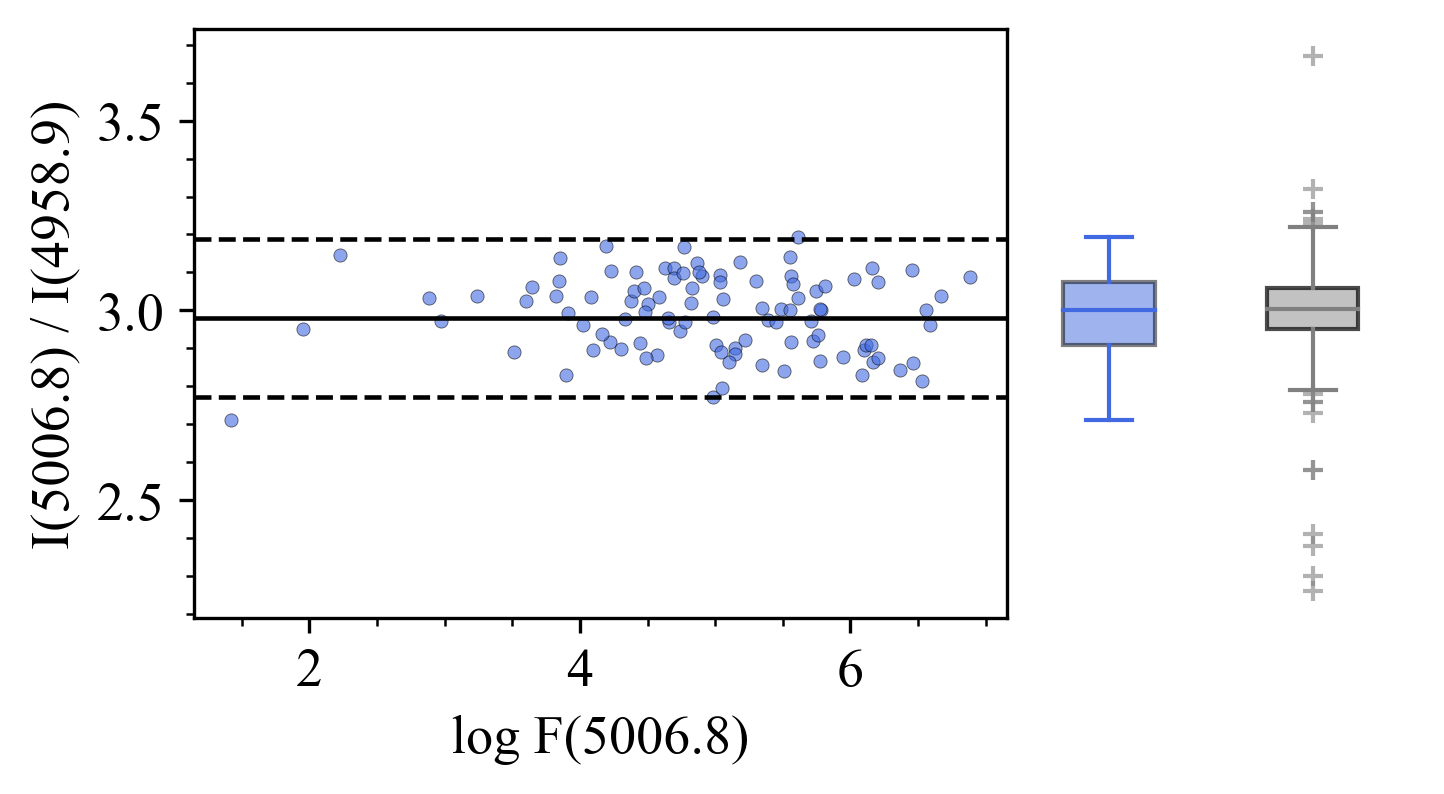}
    \caption{Comparison of flux ratios of our measured de-reddened [O~{\sc iii}] lines with theoretical 
    values with their line ratios plotted against the [O~{\sc iii}]~$\lambda$5007 flux in $10^{-16} 
    \mathrm{erg~cm^{-2}s^{-1}}$ on a log scale. The black line is the theoretical line ratio value of 
    2.98 while the dashed lines indicate 7\% deviation for typical uncertainties. The right shows box 
    plots that compare our results (blue) with 
those for 124 Galactic PNe from \cite{Rodriguez2020} (grey). The boxes depict the 25th to 75th 
percentiles, with the median indicated by a horizontal line. The whiskers extend to the 10th and 90th 
percentile, while outliers are represented by individual crosses. The results show excellent agreement to 
within 7\% errors and no bias as a function of flux.}
    \label{[OIII]}
    \end{center}
  \end{figure}

We have obtained well determined chemical abundances for 124 PNe in the Galactic bulge 
from our VLT data together with logarithmic extinctions and plasma diagnostics of $n_{\mathrm{e}}$ and $T_{\mathrm{e}}$. Prior to our work there were about $\sim240$ bulge PNe with chemical abundances 
determined over the last 50 years and of very variable quality. For 34 we have now determined abundances 
for the first time which alone adds $\sim$14\% to the available sample of bulge PNe abundances. 
We have significantly improved the accuracy and reliability of the rest. Interstellar 
reddening, physical conditions (e.g. electron densities, $n_{\mathrm{e}}$, temperatures, 
$T_{\mathrm{e}}$), and chemical compositions are derived for each PN using a variety of line diagnostics. 
Comparisons with the best available literature values for 75 PNe in common shows that our significant new 
data are reliable, robust and internally self-consistent. They represent the largest  well 
understood and high quality compilation of PNe abundances currently available. Our VLT 
low-to-medium resolution spectroscopic observations exhibit excellent consistency with 
previous high-resolution spectra with 2-m class telescope, demonstrating the reliability 
of weak line detection, the line deblending methods and our line flux measurement, 
particularly for weak recombination lines. Our abundance compilation, which adopted 
updated atomic data and ionisation correction factor (ICF) schemes introduced 
in \citep{DMS14}, results in overall 
higher abundances of alpha elements than solar compared to the general abundance pattern 
in the literature. The lockstep behavior of alpha elements, which was less evident 
in previous studies, is clearly observed in our results. In Figure~\ref{[OIII]} we 
plot [O~{\sc iii}] line ratios for 100 PNe from our sample. It is clear our data consistency 
provides solid confidence in the reliability and quality of our data and its reduction 
and hence for our abundances.

\section{The PNe sulfur anomaly project}
\begin{figure}[t]
    \centering
    \includegraphics[width=0.6\textwidth]{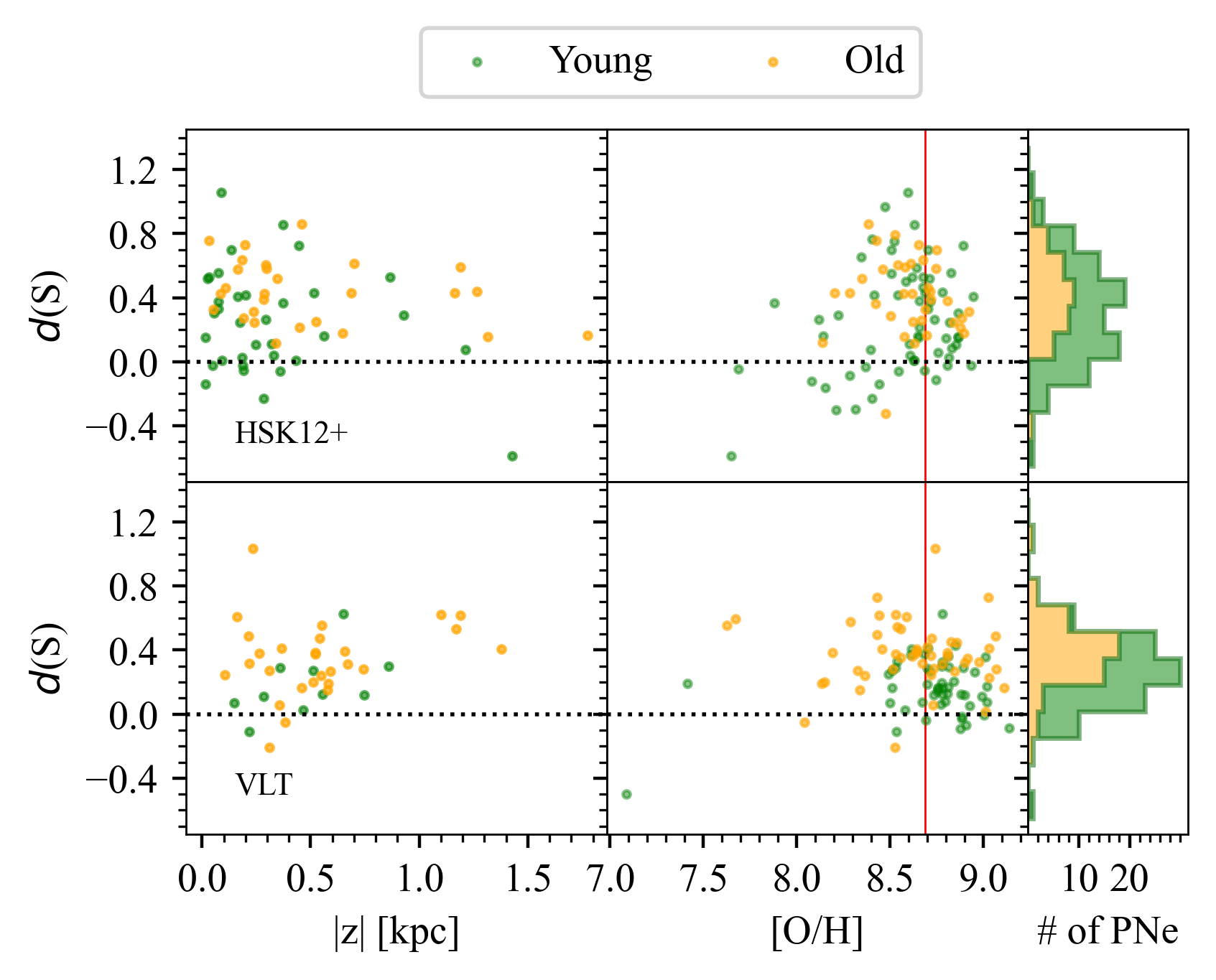}
    \caption{A combined figure for sulfur deficit $d$(S) on the vertical axis for the HSK12+ sample (top 
    row) and VLT sample (bottom row) against i) absolute height above the Galactic plane `$|z|$' in kpc, 
    ii) [O/H] and iii) stacked histograms of $d$(S) values with bin size 0.15~dex. The Galactic disk and 
    bulge PNe samples are color-coded to their basic age group from criteria outlined in Paper~V with 
    green indicating young and orange old progenitor stars. The middle panel red vertical line indicates 
    solar oxygen abundance.}  
    %The young disk population has lower Galactic heights (as expected) and generally lower $d$(S) 
    %values.}
    \label{anomaly}
\end{figure}

We have undertaken a detailed investigation of the so-called ``sulfur anomaly" (Tan \& Parker, 
submitted). This is a long-standing puzzle that has perplexed astronomers for decades. The  anomaly for 
PNe was first identified by \citep{Henry2004}. Sulfur is expected to be in lockstep with 
other $\alpha$ elements like oxygen, neon, argon, and chlorine in more massive stars, 
so its cosmic abundance should also be proportional. Strong correlations of [S/H] and [O/H] 
are seen in H~II regions and blue compact galaxies. 
However, PNe sulfur abundances, which arise from low-to-intermediate mass progenitors, 
have consistently been lower, giving rise to the sulfur anomaly. ICFs have suggested to be a potential 
suspect \citep{Reichel2022}.

Our analysis uses our robust VLT based bulge PNe chemical abundances. With our significantly improved accuracy and reduced abundance uncertainties and by using argon as a superior PNe metallicity indicator, we show that the sulfur anomaly is reduced and better constrained. Furthermore, for the first time, we demonstrate sulfur $\alpha$-element lock-step behaviour in PNe with both oxygen and argon.  Our work dispels earlier hypotheses that the anomaly originates from underestimation of higher sulfur ionization stages. Using a machine learning approach, we reveal that earlier ICF schemes contributed to the anomaly. We also identify a correlation between the sulfur anomaly and the age/mass of PNe progenitors, with the anomaly being either absent or significantly reduced in PNe with young progenitors (refer Figure~\ref{anomaly}). Despite inherent challenges and uncertainties, we link this relationship to PNe dust chemistry, noting those with carbon-dust chemistry exhibit a more pronounced anomaly. By integrating these findings, we provide a plausible explanation for the residual sulfur anomaly and propose its potential as an indicator of relative age compositions in galaxies based on their PNe observations.

\section{Summary}
We have used a well defined sample of 136 PNe carefully selected to be members of the Galactic Bulge and consistently observed with the same instrumentation and set-up on the 8m ESO VLT. The resulting high-quality data has been used to undertake a series of investigations covering morphological distribution, central star identification and kinematic ages \citep{Tan2023a}, major axes orientation \citep{Tan2023b},  abundance determinations, Tan et al. (in press) and the sulfur anomaly (Tan \& Parker, 2023, submitted). These studies have resulted in some important findings:

\begin{itemize}
    \item 68\% of our bulge sample are bipolars. This is significant for a population ostensibly dominated by old stars. 
%    \item 
If bipolars emerge from higher mass progenitors, and are in binaries there has to be a young stellar bulge population feeding the observed population we see today.
    \item 15\% of \emph{Gaia} CSPN identified by \citep{CW2021} are incorrect.
    \item We provide the most complete and accurate kinematic age estimates for our 
compact bulge PNe ever compiled 
\item The expected young age of the compact bulge PNe sample is confirmed 
with an average kinematic age of 6,400~$\pm$~3,800 years. 
\item Kinematically older 
PNe are dominated by bipolars with 14/16 PNe with estimated ages greater than 10,000 years 
being bipolars
\item We have found a 5$\sigma$ major axes alignment for bulge PNe that host short period binaries. 
Almost all are bipolar. The bulge PNe alignment signal reported previously was diluted
\item As a result there has to have been  an organised, long lived 
process at play at the centre of our Galaxy that has to 
have endured over cosmologically significant time periods given several 
PNe in our sample  emerge from stars of different progenitor mass
\item We have obtained well determined chemical abundances for 124 PNe in the Galactic bulge to produce the largest self-consistent well understood and high quality compilation of PNe abundances currently available
\item We show sulfur $\alpha$-element lock-step in PNe for oxygen and argon for the first time.  
\item We dispel the hypotheses that the sulfur anomaly originates from underestimation of higher sulfur ionization stages 
\item Using a machine learning approach, we reveal that earlier ICF schemes contributed to the anomaly.  The sulfur anomaly correlates with the age/mass of PNe progenitors, being  absent or significantly reduced in PNe with young progenitors
\end{itemize}

\end{document}